\documentclass{pasj00}

\newcommand{\lsim}{\raisebox{0.3mm}{\em $\, <$} \hspace{-3.3mm}
\raisebox{-1.8mm}{\em $\sim \,$}}
\newcommand{\gsim}{\raisebox{0.3mm}{\em $\, >$} \hspace{-3.3mm}
\raisebox{-1.8mm}{\em $\sim \,$}}
\newcommand{\bm}[1]{\mbox{\boldmath $#1$}}

\begin{document}
\SetRunningHead{Ohsuga}{Supercritical Disk Accretion onto NSs}
\Received{2007/03/07}
\Accepted{2007/06/15}

\title{Two-dimensional Radiation-hydrodynamic Model
for Supercritical Disk Accretion Flows onto Neutron Stars}

\author{Ken \textsc{Ohsuga} \altaffilmark{1,2}} 
\altaffiltext{1}{Department of Physics, Rikkyo University, 
3-34-1 Nishi-Ikebukuro, Toshima-ku, Tokyo 171-8501}
\altaffiltext{2}{Institute of Physical and Chemical Research (RIKEN), 
2-1 Hirosawa, Wako, Saitama, 351-0198}


\KeyWords{accretion, accretion disks --- 
ISM: jets and outflows --- hydrodynamics --- stars: neutron 
--- stars: individual (SS433)} 

\maketitle


\begin{abstract}
We performed two-dimensional radiation hydrodynamic simulations
of supercritical accretion flows around neutron stars (NSs).
In contrast with the accretion flows onto black holes (BHs),
we find that the shell-shaped high-density regions form around the NSs,
since the radiation force is enhanced in the innermost regions.
The enhanced radiation force drives 
strong outflows above and below the disk.
The mass-accretion rate onto the NS
exceeds the critical rate, $L_{\rm E}/c^2$,
with $L_{\rm E}$ being the Eddington luminosity.
However it is about $20-30\%$ of that onto the BH,
under the condition that we employ the same mass-input rate,
$\dot{M}_{\rm input}$,
which is mass injected from the outer disk boundary per unit time.
The mass-outflow rate is a few-times
larger in flows around NSs than in flows around BHs.
The supercritical NS accretion flows 
mainly release the accretion energy as the kinetic energy 
of the outflows,
though the disk luminosity is predominant over 
the kinetic energy output rate in the BH accretion flows.
The resulting velocity and mass-outflow rate of the outflows
are $0.2-0.3c$ and $150-700L_{\rm E}/c^2$,
respectively,
for the mass-input rate of 
$3\times 10^2\lsim \dot{M}_{\rm input}
/(L_{\rm E}/c^2)\lsim 3\times 10^3$.
This implies that 
the SS433 jets can be roughly explained by the 
supercritical accretion onto a NS.
However, the collimation angle of the outflows in our simulations
($\sim 20^\circ$)
is larger than that of the SS433 jets (a few degrees).

\end{abstract}



\section{INTRODUCTION}
Luminous compact objects are thought to be 
powered by accretion flows.
It is widely believed that 
accretion processes can be basically described 
by the standard disk model,
proposed by \citet{SS73}.
However, this model breaks down
if the mass-accretion rate is comparable to,
or exceeds, the critical rate, $L_{\rm E}/c^2$,
where $L_{\rm E}$ is the Eddington luminosity.
Then, advective cooling is predominant over radiative cooling.

\citet{Abramowicz88} established the slim-disk model,
which is one-dimensional model that includes the advective cooling.
Two-dimensional radiation-hydrodynamic (RHD) simulations 
of supercritical accretion flows around black holes (BHs)
were attempted by \citet{ECK88} and \citet{Okuda02}.
By performing long-term two-dimensional RHD simulations,
we have for the first time 
confirmed the occurrence of supercritical disk accretion 
onto BHs \citep[hereafter Paper I]{O05}.

In the case of the supercritical disk accretion flows around BHs,
a large amount of photons generated inside the disk 
are swallowed by the BH with accreting matter.
This effect largely attenuates the radiation force,
assisting the occurrence of inflow motion.
However, 
the situation considerably
differs in disks around neutron stars (NSs).
In this case, the radiation energy is not swallowed by the NS,
and the energy of accreting matter would be converted 
to radiation energy 
at the NS surface.
Thus, the radiation force is enhanced and works to 
prevent the inflow motion,
although the matter is accelerated inward 
by the radiation force in cooperation with the gravity 
at the very vicinity of the BH.

The X-ray source SS433 exhibits powerful relativistic jets.
The observations of Doppler-shifted iron lines
revealed that the mass-outflow rate of the jets 
highly exceeds the critical rate for the NS/stellar BH 
\citep{Kotani96}.
Thus, SS433 is thought to undergo supercritical accretion flows,
although its central object is still not understood.
Whereas some researchers have suggested that a BH exists in SS433,
a NS hypothesis is not denied.
On the other hand,
by the X-ray observations of low-mass X-ray binaries (LMXBs),
\citet{TM06} recently discovered
a sign of supercritical accretion onto the NSs.
They reported that the spectra of LMXBs consist of 
three components in the supercritical phase.
Two components are thought to be emission of
the disk and the NS surface,
which are observed 
when the X-ray luminosity is much smaller than the Eddington luminosity
as well.
The third component appears only in the supercritical phase,
and is thus presumably the emission 
of outflows driven by the strong radiation force
of the supercritical accretion disks.

The super or near-critical accretion onto NS has been studied 
in spherical geometry
by analytic method and numerical calculations 
(\authorcite{BK80} \yearcite{BK80}, \yearcite{BK83};
\authorcite{Klein80} \yearcite{Klein80};
\authorcite{Miller90} \yearcite{Miller90}).
They reported the maximum accretion rate to be about 
critical late and the maximum luminosity to be near the Eddington luminosity.
However, \citet{Demmel90} proposed a inhomogeneous spherical accretion 
model, by which the luminosity can exceed the Eddington luminosity
for large mass-accretion rate.
The maximum luminosity of supercritical disks 
around a magnetized NS was suggested to be $46L_{\rm E}$
by \citet{Lipunov82}.
Two-dimensional RHD simulations 
of supercritical accretion flows
around NSs were attempted by \citet{Okuda00}.
However, though the matter was injected at the supercritical rate
from the outer boundary,
most of the matter was blown off from the system as the jets 
and the disk wind.
The steady structure and physical properties of 
supercritical disk accretion onto a NS is still open issue.


In this paper,
we consider the supercritical disk accretion flows around 
non-magnetic and non-rotating NSs,
based on long-term two-dimensional RHD simulations.
We also compare them with supercritical BH accretion flows.
In \S 2, our model and numerical method are described. 
We present the numerical results in \S 3.
\S 4 and \S5 are devoted to discussion and conclusions.

\section{MODEL AND METHOD}
\subsection{Basic Equations}
We describe the RHD equations including the viscosity term.
We use spherical polar coordinates $(r, \theta, \varphi)$,
where $r$ is the radial distance, 
$\theta$ is the polar angle,
and $\varphi$ is the azimuthal angle.
The origin is set at the central object (NS otherwise BH).
In the present study,
we assume axisymmetry with respect 
to the rotation axis (i.e., $\partial/\partial \varphi=0$),
as well as reflection symmetry
relative to the equatorial plane (with $\theta = \pi/2$).
We describe the gravitational field of the central object
in terms of pseudo-Newtonian hydrodynamics,
in which the gravitational potential
is given by $-GM/(r-r_{\rm S})$ as was 
introduced by \citet{PW80},
where $M$ is the mass of the central object
and $r_{\rm S}=2GM/c^2$ is the Schwarzschild radius.
The flow is assumed to be non-self-gravitating.
The basic equations are the continuity equation,
\begin{equation}
  \frac{\partial \rho}{\partial t}
  + \nabla \cdot (\rho {\bm v}) = 0,
  \label{mass_con}
\end{equation}
the equations of motion,
\begin{eqnarray}
  \frac{\partial (\rho v_r)}{\partial t}
  &+& \nabla \cdot (\rho v_r {\bm v}) \nonumber\\
  =&-&\frac{\partial p}{\partial r}
  + \rho \left\{ 
    \frac{v_\theta^2}{r} + \frac{v_\varphi^2}{r}
    -\frac{GM}{(r-r_{\rm S})^2}
  \right\}
  +f_r + q_r,
  \label{mom_r}
\end{eqnarray}
\begin{equation}
  \frac{\partial (\rho r v_\theta)}{\partial t}
  + \nabla \cdot (\rho r v_\theta {\bm v}) 
  = - \frac{\partial p}{\partial \theta}
  + \rho v_\varphi^2 \cot\theta
  + r f_\theta
  + r q_\theta,
  \label{mom_th}
\end{equation}
\begin{equation}
  \frac{\partial (\rho r v_\varphi \sin\theta )}{\partial t}
  + \nabla \cdot (\rho r v_\varphi \sin\theta {\bm v}) 
  = r q_\varphi \sin\theta ,
  \label{mom_varphi}
\end{equation}
the energy equation of the gas,
\begin{equation}
  \frac{\partial e }{\partial t}
  + \nabla\cdot(e {\bm v}) 
  = -p\nabla\cdot{\bm v} -4\pi \kappa B 
  + c\kappa  E_0 + \Phi_{\rm vis},
  \label{gase}
\end{equation}
and the energy equation of the radiation,
\begin{equation}
  \frac{\partial E_0}{\partial t}
  + \nabla\cdot(E_0 {\bm v}) 
  = -\nabla\cdot{\bm F_0} -\nabla{\bm v}:{\bm {\rm P}_0}
  + 4\pi \kappa B - c\kappa E_0.
  \label{rade}
\end{equation}
Here, $\rho$ is the mass density,
$\bm{v}=(v_r, v_\theta, v_\varphi)$ is the velocity,
$p$ is the gas pressure,
$e$ is the internal energy density of the gas,
$B$ is the blackbody intensity,
$E_0$ is the radiation energy density,
${\bm F}_0$ is the radiative flux,
${\bm {\rm P}}_0$ is the radiation pressure tensor,
$\kappa$ is the absorption opacity,
${\bm q}=(q_r, q_\theta, q_\varphi)$ is the viscous force,
and $\Phi_{\rm vis}$ is the viscous dissipative function,
respectively.
The radiation force ${\bm f}_{\rm rad} = (f_r, f_\theta)$ 
is given by 
\begin{equation}
  {\bm f}_{\rm rad} = \frac {\chi}{c} {\bm F}_0,
  \label{frad}
\end{equation}
where $\chi (= \kappa+\rho \sigma_{\rm T}/m_{\rm p})$ 
is the total opacity 
with $\sigma_{\rm T}$ being the Thomson-scattering cross-section
and $m_{\rm p}$ being the proton mass.

As the equation of state, we use
\begin{equation}
  p=(\gamma-1)e,
\end{equation}
where $\gamma$ is the specific heat ratio.
The temperature of the gas, $T$, can then be calculated from
\begin{equation}
  p=\frac{\rho k T}{\mu m_{\rm p}},
\end{equation}
where $k$ is the Boltzmann constant and
$\mu$ is the mean molecular weight.

The radiative flux and the radiation stress tensor are
solved under the flux-limited diffusion
approximation,
so that they are expressed 
in terms of the radiation energy density
\citep{LP81}.
It gives correct relations in the optically thick diffusion limit
and optically thin streaming limit, respectively.

We assume that the only $r\varphi$-component 
of the viscous stress tensor, 
which plays important roles for
the transport of the angular momentum
and heating of the disk plasma,
is non zero,
\begin{equation}
  \tau_{r\varphi}=\eta r \frac{\partial}{\partial r}
  \left(\frac{v_\varphi}{r}\right),
\end{equation}
where $\eta$ is the dynamical viscosity coefficient.
Then, the radial and polar components of the viscous force are null
($q_r=q_\theta=0$),
and the right hand side of equation (\ref{mom_varphi}) 
is described as
\begin{equation}
  r q_\varphi \sin\theta = \frac{1}{r^2}\frac{\partial}{\partial r}
  \left( r^3 \tau_{r\varphi} \sin\theta \right).
\end{equation}
The viscous dissipative function is given by
\begin{equation}
  \Phi_{\rm vis}=\eta 
  \left[ r \frac{\partial}{\partial r}
  \left(\frac{v_\varphi}{r}\right)
  \right]^2.
\end{equation}
The dynamical viscosity coefficient
is prescribed as
\begin{equation}
  \eta = \alpha \frac{p+\lambda E_0}{\Omega_{\rm K}},
\end{equation}
where $\alpha$ is the viscosity parameter,
and $\Omega_{\rm K}$ is the Keplerian angular speed.
Here, the parameter, $\lambda$, is given by 
$(2+{\cal R})/(6+3{\cal R}+{\cal R}^2)$, 
using the dimensionless quantity, 
${\cal R}=\left| \nabla E_0 \right| 
/ \left( \chi E_0 \right)$.
Since it becomes $1/3$ in the optically thick regime
and vanishes in the optically thin regime,
the viscous stress tensor is related to the gas pressure 
or the total pressure, $p_{\rm total}$, as,
\begin{equation}
  \tau_{r\varphi} \propto \left\{
    \begin{array}{ll}
      \alpha p_{\rm total} & \mbox{(optically thick limit)} \\
      \alpha p & \mbox{(optically thin limit)}
    \end{array}
  \right..
\end{equation}
It is modified $\alpha$ prescription of the viscosity,
which was proposed by Shakura \& Sunyaev (1973). 

\subsection{Numerical Method}
We numerically solved the set of RHD equations 
using an explicit-implicit finite-difference scheme 
on the Eulerian grids.
Our methods were similar to those of Paper I,
but we used the explicit scheme for solving the radiative flux term
in equation (\ref{rade}),
and employ constant time step \citep{O06}.
The computational domain consisted of spherical shells of 
$2.4r_{\rm S} \leq r \leq 500r_{\rm S}$ and $0 \leq \theta \leq 0.5\pi$,
which were divided into $96\times 96$ grid cells.

We employed two types of inner-boundary conditions
for the accretion flows around the NSs (NS model) and the BHs (BH model).
The radial component of the radiative flux is given by
\begin{equation}
 F_0^r=-cE_0,
  \label{F_BH}
\end{equation}
at the inner boundary for the BH model.
This means that 
the radiative flux term always contribute to transport
of the radiation energy onto the BH ($F_0^r<0$).
In addition,
because of the inflow motion at the inner boundary ($v_r<0$),
the radiation energy and the gas energy are 
swallowed by the BH via advection,
$v_rE_0<0$ and $(e+\rho v^2/2)v_r<0$,
in our simulations.
In other words, 
the BH always swallows both the matter and the energy.
In contrast,
we assumed that the NS cannot swallow the energy.
The gas energy advected onto the NS surface
is supposed to be immediately released as the radiation.
Hence, we set 
\begin{equation}
 F_0^r=-ev_r-\frac{1}{2}\rho v^2 v_r,
  \label{F_NS}
\end{equation}
at the inner boundary for the NS model.
Our simulations gave negative radial velocity at the inner boundary
($v_r<0$),
leading an outward radiative flux ($F_0^r>0$) 
at the NS surface.
When we updated the radiation energy density by 
solving the equation (\ref{rade}),
the advective term of the radiation energy was numerically set to be null
at the inner boundary, $v_rE_0=0$,
whereas the radial velocity and the radiation energy density
are both non zero there.
To sum up, the matter accretes onto the NS surface,
but the energies of radiation and gas 
were not swallowed by the NS in our simulations.

We employed a no-torque condition at the surface of the BH.
Thus, the viscosity does not work at the inner boundary
in the case of the BH accretion flows.
In contrast, we assumed that the viscosity works 
even at the very vicinity of the inner boundary in the NS model.
Therefore, the rotation energy of the gas is effectively dissipated 
and converted to radiation energy near the NS surface.
This viscous effect at the inner boundary 
would depend on the rotation velocity of the NS.
However, we note that our results do not change so much
unless the rotation velocity of the NS surface
is comparable to the Keplerian velocity.

The radiation energy density increases near the NS surface
because of such inner-boundary conditions,
leading the enhanced radiation force.
The radiation force counteracts with the gravity
and works to prevent the inflow motion.
In contrast, the matter is accelerated inward
by the radiation force in cooperation with the gravity
in the case of the BH model.

We started the calculations with a hot, rarefied, and
optically thin atmosphere. 
There was no cold dense disk in the computational domain, initially.
Matter was continuously injected into the computational 
domain through the outer disk boundary 
($r = 500r_{\rm S}$, $0.45\pi \leq\theta\leq 0.5\pi$). 
Therefore, we could avoid any influence of the 
initial configuration on the final result, 
although a long integration time is required. 
The injected matter is supposed to have a
specific angular momentum corresponding to the 
Keplerian angular momentum at $r = 100r_{\rm S}$.
We set the mass-input rate, $\dot{M}_{\rm input}$,
which is the mass injected into the computational domain per unit time,
so as to be constant at the outer disk boundary.
At the outer boundary region above the outer disk boundary
($r = 500r_{\rm S}$, $\theta<0.45\pi$),
we used free boundary conditions and allowed for matter to go out 
but not to come in.


Throughout the present study, 
we assume $\alpha=0.1$, $\gamma=5/3$, $\mu=0.5$,
and $M=1.4M_\odot$.
Thus, the radius of the inner boundary, $2.4r_{\rm S}$,
is nearly $10 \rm km$.
Such mass and the radius are consistent with
recent models of a NS,
although they slightly depends on the equation of state
inside the NS.
Since we aim to understand any
differences of the accretion flows around NSs and BHs,
the mass of the BH was also set to be $1.4M_\odot$.
For the absorption opacity,
we considered free-free absorption 
and bound-free absorption for solar metallicity
\citep{HHS62, RL79}

\section{RESULTS}
\subsection{Quasi-steady Structure}
\label{structure}
\begin{figure*}
  \begin{center}
    \FigureFile(130mm,130mm){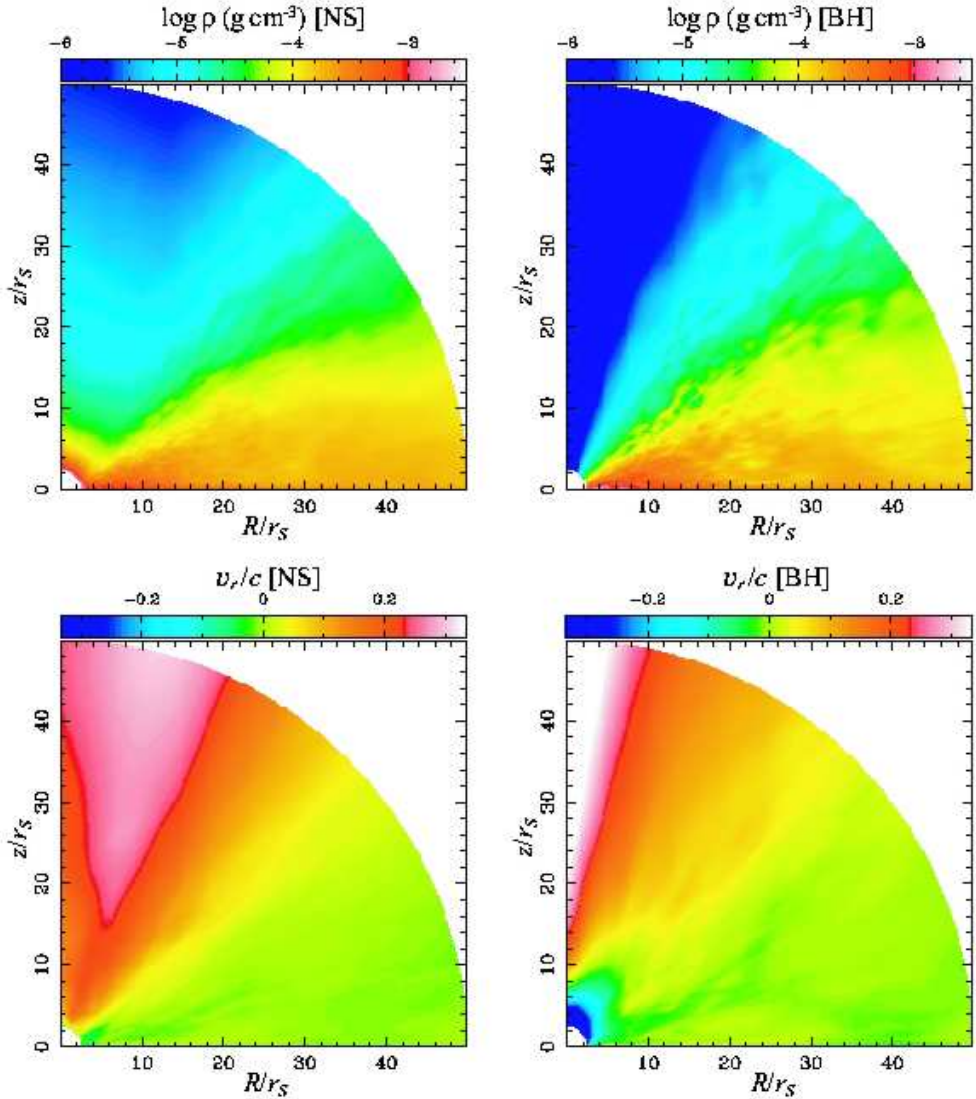}
  \end{center}
 \caption{
 Time-averaged distributions of the density (the upper-left panel) and 
 the radial velocity normalized by the speed of light 
 (the lower-left panel) on the $R$-$z$ plane for the NS model.
 Those for the BH model are shown in the upper-right and lower-right panels.
 The adopted parameter is $\dot{M}_{\rm input}=10^3L_{\rm E}/c^2$.
 \label{2D}
 }
\end{figure*}

We first give a brief overview of the time evolution 
of our simulations.
The overall evolution of the NS accretion flows is 
similar to that of the BH accretion flows (see Paper I in detail).
It is divided into two distinct phases:
an accumulation phase and a quasi-steady phase.
Because the injected matter has a small specific angular momentum, 
it firstly free falls, but accumulates 
around regions with a radius of $r \sim 100 r_{\rm S}$,
forming a rotating disk (accumulation phase).
Eventually, the viscosity starts to work.
The angular momentum of the gas is transported outward,
driving the inflow motion (quasi-steady phase).
Radiatively driven outflow is generated 
around the rotation axis in this phase.
The critical time separating these two phases
roughly coincides with the viscous timescale.

Figure \ref{2D} indicates time-averaged contours 
of the density in the quasi-steady state 
for the NS model (the upper-left panel) 
and the BH model (the upper-right panel),
where we adopt $\dot{M}_{\rm input}=10^3L_{\rm E}/c^2$.
As shown in these panels,
a high-density region appears 
around the equatorial plane ({\it yellow}) 
and a rarefied region forms near the rotation axis ({\it blue})
in both models.
The high-density regions indicate
the radiation pressure-supported disks.
The radiation force drives the outflow 
in the less-dense region above the disk.
We find a significant difference 
in the very vicinity of the inner boundary.
Although the rarefied region
connects to the inner boundary 
along the rotation axis in the BH model,
the high-density region ({\it orange}) 
surrounds the NS.

Such a shell-shaped structure is built up 
by the radiation force, which counteracts with the gravity.
Since the radiation energy, as well as the gas energy,
is not swallowed by the NS in our simulations,
the radiation energy increases near the inner boundary.
Hence, the radial component of the radiative flux
becomes positive (outward flux) there, so that
the radiation force works to prevent the inflow motion
(discussed later).
Here, we note that such a difference in the density distribution 
would affect the emergent spectra and its viewing-angle dependence.
We discuss this point in \S\ref{S_Teff}.

In the lower-left (NS model) and lower-right (BH model) panels,
we display the radial velocity normalized by the speed of light.
It is found that the gas slowly falls inward in the dense disk regions.
On the other hand, 
the matter is accelerated outward in the rarefied
region above the disk,
except for the very vicinity of the BH,
where the gas falls onto the BH.
We find in the NS model
that the typical velocity of the outflow is $0.2-0.3c$
and its collimation angle is $\sim 20^\circ$.
The BH model exhibits more collimated high velocity regions 
along the rotation axis.
The supercritical accretion onto the NS forms
more powerful outflows than does that around the BH.
Both the mass-outflow rate and the kinetic energy output rate
by the outflow are larger in the NS model than in the BH model
(discussed later).

\begin{figure}
  \begin{center}
    \FigureFile(85mm,85mm){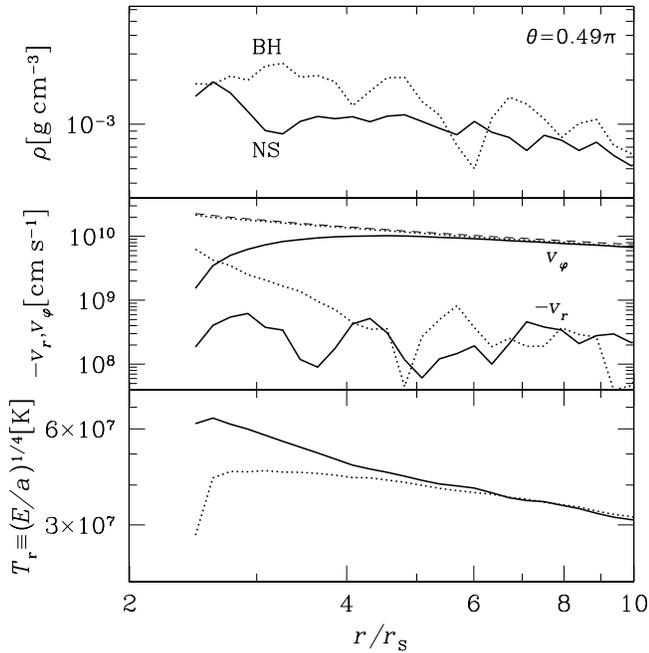}
  \end{center}
 \caption{
 Time-averaged one-dimensional profiles of the density (the top panel), 
 the radial and rotation velocities (the middle panel),
 and the radiation temperature (the bottom panel)
 near the equatorial plane, $\theta=0.49\pi$.
 The solid and dotted lines indicate the results 
 of the NS model and the BH model, respectively.
 The Keplerian velocity is represented by the dashed line
 in the middle panel,
 which almost overlaps with upper dotted line.
 The adopted parameter is $\dot{M}_{\rm input}=10^3L_{\rm E}/c^2$.
 \label{1D}
 }
\end{figure}

In Figure \ref{1D} 
we represent one-dimensional distributions of
the gas density (the top panel), the inflow and rotation velocities 
(the middle panel),
and the radiation temperature (the bottom panel) near the equatorial plane,
$\theta=0.49\pi$.
Here, the solid and dotted lines indicate the results of the NS model 
and the BH model, respectively.
The radiation temperature is defined as 
$T_r \equiv (E_0/a)^{1/4}$,
where $a$ is the radiation constant.
As shown in the top panel,
the gas density in the NS model
is comparable to, or slightly smaller 
than, that in the BH model.
It is found in the middle and bottom panels 
that there are significant differences in the profiles 
of the velocities and the radiation temperature
between two models.
These are induced by the differences in the inner-boundary conditions.

In the present simulations,
the viscosity worked even at the inner boundary for the NS model,
whereas we used no-torque condition at the surface of the BH.
Therefore,
the rotation velocity deviates from the Keplerian velocity
as the flow approaches the NS,
though the gas rotates at about the Keplerian velocity 
in the regions of $r\gsim 5r_{\rm S}$ (see the middle panel).
In contrast, the rotation velocity of the BH model
roughly agrees with 
the Keplerian velocity 
even in the vicinity of the inner boundary.

We find in the bottom panel that the radiation energy 
is larger in the NS model than in the BH model
near the inner boundary.
Near the surface of the NS,
the viscosity effectively converts the rotation energy 
into the internal energy of the gas.
Subsequently, the gas-radiation interaction
turns such energy into the radiation energy.
In addition, the energy of the gas falling onto the NS
is converted to radiation energy at the surface of the NS
(see equation [\ref{F_NS}]).
Due to the combination of these effects, 
the radiation energy increases around the inner boundary
in the NS model.
In the case of the BH accretion flow, 
the no-torque condition is employed and the energy is swallowed by the BH,
attenuating the radiation energy density around the inner boundary.

We find in the middle panel
that the resulting inflow velocity, $-v_r$, in the NS model
is much smaller than that in the BH accretion flows
around the inner boundary.
The radiation force
is enhanced in the vicinity of the NS surface
via the large radiation energy and its steep profile
(see the bottom panel).
Thus, the radiation force works to decelerate the inflow motion.
That radiative deceleration in the supercritical flows
around the NS was mentioned by \authorcite{BK80} (1980; 1983).
In the BH model,
the radiative deceleration is relatively inefficient,
since the flatter profile of the radiation energy density
leads to a small radiative flux 
(see the bottom panel).
In addition, the radiation force accelerates the gas inward 
in cooperation with the gravity at the surface of the BH,
since the radial component of the radiative flux is negative
(inward flux).
As a result, the inflow velocity increases as the flow approaches the BH.

\subsection{Mass Accretion Rate and Mass Outflow Rate}
\label{mdot}
\begin{figure}
  \begin{center}
    \FigureFile(85mm,85mm){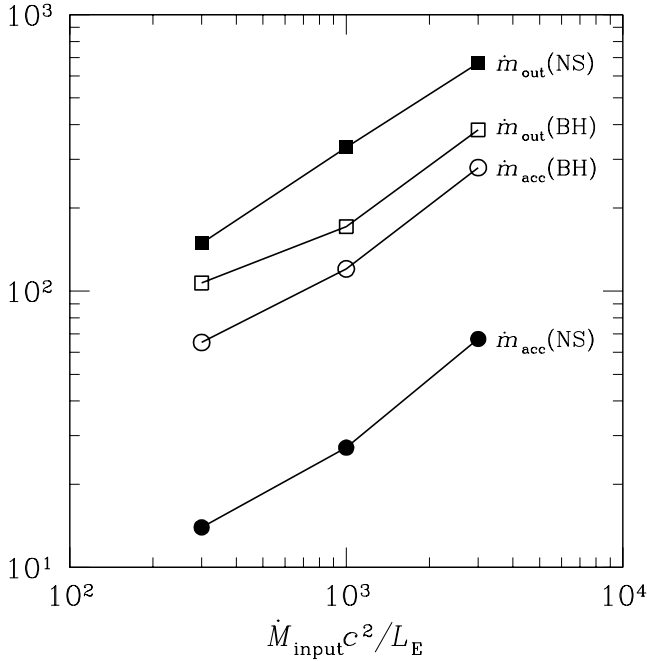}
  \end{center}
 \caption{
 Mass-accretion rate onto the NS/BH 
 and mass-output rate by the high velocity outflows
 normalized by the critical rate,
 $\dot{m}_{\rm acc}$ ({\it circles}) 
 and $\dot{m}_{\rm out}$ ({\it squares}),
 as functions of normalized mass-input rate.
 The filled and open symbols indicate the results of the NS model
 and the BH model, respectively.
 \label{mass}
 }
\end{figure}

We plot in Figure \ref{mass}
the mass-accretion rate onto the NS/BH ({\it circles}) 
and the mass-outflow rate ({\it squares})
normalized by the critical accretion rate
as functions of the mass-input rate.
Here, the mass-outflow rate is the mass 
ejected through the outer boundary per unit time
by the high-velocity outflow 
whose radial velocity exceeds the escape velocity.
The filled and open symbols indicate the results of the NS and BH models,
respectively.

We find that 
both the mass-accretion rate and the mass-outflow rate increase
with an increase of the mass-input rate.
We find that 
the supercritical accretion is not a peculiar phenomenon 
in BH accretion flows.
The mass-accretion rate of the NS model highly exceeds 
the critical accretion rate,
although it is as small as $20-30\%$ for the BH model.
Especially surprising is that 
the mass-outflow rate for the NS model
exceeds that in the case of BH accretion flows,
although the mass-accretion rate is much smaller 
in the NS model than in the BH model.
The supercritical disk around NS blows away
a few times as much matter as outflowing matter 
of the disk around the BH.
The supercritical accretion flows around the NSs
effectively produce the radiatively driven outflow.
We discuss this point again in the next subsection.

\subsection{Luminosity and Kinetic Energy Output Rate}
\label{Ene}
\begin{figure}
  \begin{center}
    \FigureFile(85mm,85mm){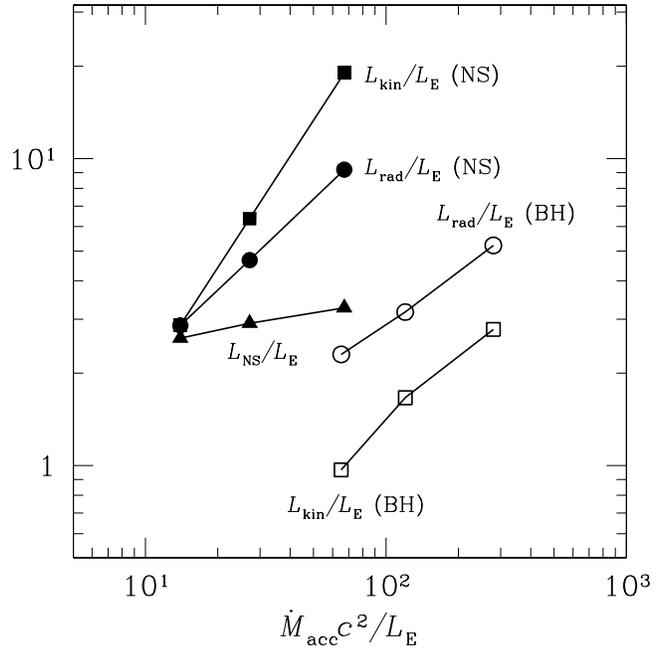}
  \end{center}
 \caption{
 Luminosity ({\it circles}) and kinetic energy output rate 
 ({\it squares}) 
 as functions of normalized mass-accretion rate 
 onto the central object, $\dot{M}_{\rm acc}c^2/L_{\rm E}$,
 for $\dot{M}_{\rm input}c^2/L_{\rm E}=
 3\times 10^2, 10^3, 3\times 10^3$
 from left to right.

 The filled and open symbols indicate the results of the NS model
 and the BH model, respectively.
 The luminosity of the NS is represented by the filled triangles.
 \label{Lum}
 }
\end{figure}

In Figure \ref{Lum}
we represent the luminosity and the kinetic energy output rate
as functions of the mass-accretion rate onto the NS/BH,
$\dot{M}_{\rm acc}$.
The filled (open) circles and squares indicate
the luminosity, $L_{\rm rad}$,
and the kinetic energy output rate, $L_{\rm kin}$,
for NS (BH) model.
Here, the luminosity is evaluated
by integrating the radiative flux at the outer boundary.
The kinetic energy output rate is the kinetic energy
ejected through the outer boundary per unit time
by the high-velocity outflow.
The luminosity of the NS, $L_{\rm NS}$, 
which is evaluated by integrating
the radiative flux at the inner boundary (see equation [\ref{F_NS}]),
is represented by the triangles.
The mass-input rate is set to be 
$\dot{M}_{\rm input}/(L_{\rm E}/c^2)=3\times 10^2$,
$10^3$, and $3\times 10^3$
from left to right.

We find in this figure that 
the luminosity and the kinetic energy output rate 
increase with an increase of the mass-accretion rate
in both models.
It is also found that 
the energy conversion efficiency,
$(L_{\rm rad}+L_{\rm kin})/\dot{M}_{\rm acc}c^2$,
of the NS accretion flows
is by one order of magnitude larger 
than that of the BH accretion flows.
For instance, 
when the mass-accretion rate is around $70L_{\rm E}/c^2$,
we find $L_{\rm rad}\sim 9L_{\rm E}$ and
$L_{\rm kin}\sim 19L_{\rm E}$ for the NS model
and $L_{\rm rad}\sim 2L_{\rm E}$ and
$L_{\rm kin}\sim L_{\rm E}$ for the BH model.
The sum of the luminosity and 
the kinetic energy output rate 
almost equals the accretion energy for the NS model,
$L_{\rm rad}+L_{\rm kin}\sim 
GM\dot{M}_{\rm acc}/(2.4r_{\rm S}-r_{\rm S})$.
In contrast, we find 
$L_{\rm rad}+L_{\rm kin}
\ll GM\dot{M}_{\rm acc}/(2.4r_{\rm S}-r_{\rm S})$
in the BH model.
This is natural since, unlike a BH,  
a NS neither swallow the radiation energy nor the gas energy
in the present study.

In Figure \ref{Lum} 
we find $L_{\rm rad}>L_{\rm kin}$ for the BH model,
but $L_{\rm rad}<L_{\rm kin}$ for the NS model.
The radiation force is enhanced 
via the inner-boundary conditions for the NS model,
whereby the rotation energy is converted to radiation energy
via the viscosity 
and the energy of the inflowing matter is also converted 
to the radiation energy without being swallowed by the NS,
driving the strong radiatively driven outflow.

Next, we explain the reason why
the mass-accretion rate onto the NS 
can be over the critical rate, $L_{\rm E}/c^2$.
In the present simulations, 
the luminosity of the NS corresponds to the energy 
of the accreting gas per unit time
(see equation [\ref{F_NS}]).
Thus, if the gas sufficiently releases its energy
before reaching the inner boundary,
the gravity is predominant over the radiation force 
at the NS surface, allowing the accreting motion.
Otherwise, the radiation force would prevent the inflow motion
at the inner boundary.
Due to such mechanism,
the resulting luminosity of the NS is self-regulated
so as to meet the condition that 
the radiation force does not exceed the gravity.
Since we employ the pseudo-Newtonian potential,
the critical luminosity, in which 
the radiation force balances with the gravity, is $\sim 3L_{\rm E}$.
To conclude,
although the energy transported onto the NS surface per unit time
is limited to be $\lsim 3L_{\rm E}$,
the mass-accretion rate can greatly exceed the critical rate,
as long as the matter sufficiently loses its energy before reaching
the NS surface.
Since the BH can swallow the gas energy, as well as the radiation energy,
the mass-accretion rate can exceed the critical rate
even if the matter has a large amount of the energy.

\subsection{Effective Temperature Profile}
\label{S_Teff}

\begin{figure}[h]
  \begin{center}
    \FigureFile(85mm,85mm){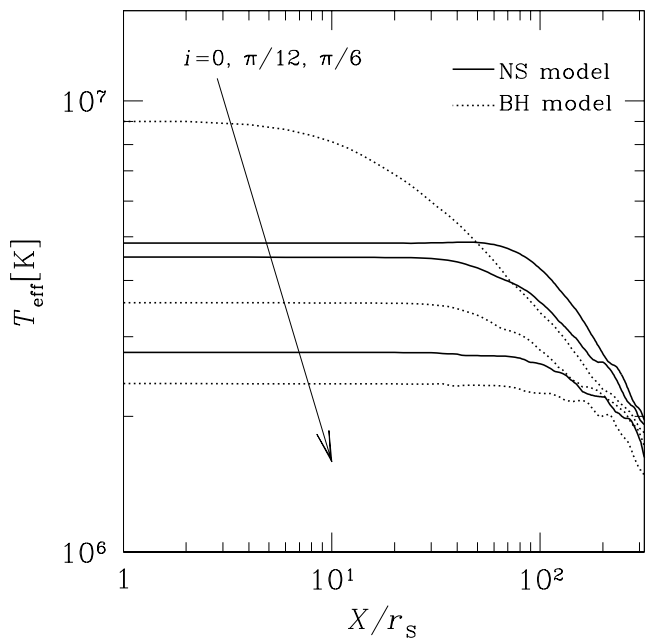}
  \end{center}
 \caption{
 Effective-temperature profiles in the $X$-directions
 for the viewing angle of $i=0$, $\pi/12$, and $\pi/6$.
 Here, $X$ is the horizontal coordinate on the observer's screen.
 The solid and dotted lines represent the results of 
 NS model and BH model, respectively.
 \label{Teff}
 }
\end{figure}

Finally, we consider the effective temperature
by solving radiation transfer,
and represent its profiles for
the inclination angles of $i=0$, $\pi/12$, and $\pi/6$
in Figure \ref{Teff}.
Here, $X$ is the horizontal coordinate on the observer's screen.
The adopted parameter is
${\dot M}_{\rm input} = 10^3 L_{\rm E}/c^2$.
Then, the luminosity is $\sim 5L_{\rm E}$ for the NS model
and $\sim 3L_{\rm E}$ for the BH model.
In this subsection, 
we focus on the effective-temperature profile
and its viewing-angle dependence.

This figure shows 
that the maximal effective temperature 
is not very sensitive to the viewing angle for the NS model.
It merely varies from $3\times 10^6$K to $5\times 10^6$K.
In contrast, it shifts from $2\times 10^6$K to $9\times 10^6$K 
for the BH model.
We also find that the accretion flow around the NS 
exhibits a flatter profile of the effective temperature,
even in the face-on view.

As shown in the Figure \ref{2D},
the density distribution
is more anisotropic in the BH model than in the NS model
(note that the outflow region above the disk is
relatively dense in the NS model,
as compared with the BH model).
In addition, a high-density shell forms only 
in the NS model.
Due to such enhanced anisotropy of the density distribution,
the emission of the BH model is more collimated 
in the polar direction than that of the NS model,
producing the steeper effective-temperature profile 
in the face-on view (see also Paper I).
In contrast, the effective-temperature profile becomes almost flat
in the NS model.

\section{DISCUSSION}
\subsection{Comparison with SS433 jets}
\label{D_SS433}
By the present simulations,
we showed that the relativistic outflows
appear above and below the supercritical disk.
Here, we compare the simulated outflows of the NS model
with the jets of SS433.

Based on the observations of the Doppler-shifted ion lines,
\citet{Kotani96} reported the mass-outflow rate of the SS433 jets
to be $>5\times 10^{-7}M_\odot\,{\rm yr}^{-1}$,
probably $2\times 10^{-6}M_\odot\,{\rm yr}^{-1}$.
Since these values correspond to $>1.7\times10^2L_{\rm E}/c^2$ and
$6.7\times 10^2L_{\rm E}/c^2$ for the mass of $1.4M_\odot$, 
our NS model can explain the mass-outflow rate of the SS433 jets,
if the mass-input rate is within the range
$3\times 10^2\lsim \dot{M}_{\rm input}
/(L_{\rm E}/c^2)\lsim 3\times 10^3$.
The typical velocity of the outflows,
$(2L_{\rm kin}/\dot{m}_{\rm out}L_{\rm E})^{1/2}c$,
is $0.2-0.3c$.
It nicely agrees with the velocity of $0.26c$ observed in SS433.

However, the collimation angle of the jets of our model
deviates from that of the SS433 jets.
The collimation angle of the SS433 jets seems to be
as small as a few degrees.
However, we find the angle of $\sim 20^\circ$
in our simulations (see the upper-left panel in Fig. \ref{2D}).
The magnetic effects might resolve this issue.
\citet{KHM04} have reported that the collimated jets
are produced via a magnetic interaction 
between a weakly magnetized NS and its accretion disk.
The jets might be collimated at the outer region,
although the size of the computational domain was restricted 
to $500 r_S$ in the present study.
We will clarify this point in future work
by performing the simulations with larger computational domains.

\begin{figure}
  \begin{center}
    \FigureFile(85mm,85mm){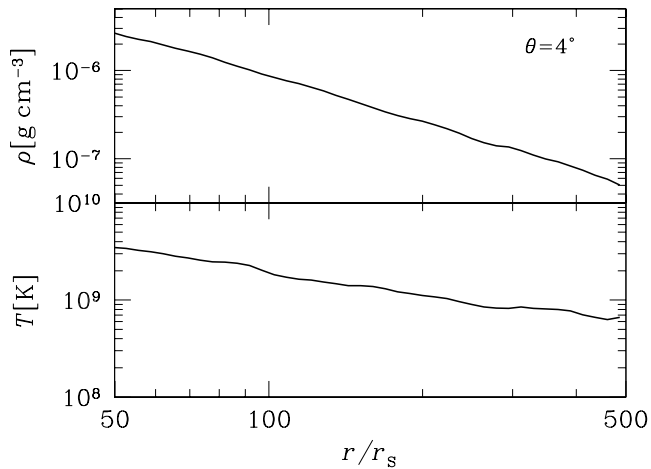}
  \end{center}
 \caption{
 Time-averaged one-dimensional profiles of the density (the top panel) 
 and the gas temperature (the bottom panel)
 in the outflow regions ($\theta=4^\circ$) for the NS model.
 \label{SS433}
 }
\end{figure}

We next discuss about the density and the temperature.
\citet{Kotani96} also estimated the density and the temperature
in the X-ray emitting region,
$r\gsim 10^9\rm cm$ 
($2\times 10^3 r_{\rm S}$ for the mass of $1.4M_\odot$),
as $\rho <10^{-8} \rm g \, cm^{-3}$
and $T \sim 3\times 10^8 \rm K$.
In Figure \ref{SS433}
we represent the resulting profiles of the density 
(the top panel) and the temperature (the bottom panel)
near the rotation axis, $\theta=4^\circ$
for $\dot{M}_{\rm input}=10^3L_{\rm E}/c^2$.
As shown in this figure,
the density and the temperature are roughly proportional to
$r^{-1}$ and $r^{-2/3}$, respectively.
Thus, our results are roughly consistent with the observations
as long as the slope of the profiles does not drastically change 
up to the regions of $r\sim 2\times 10^3 r_{\rm S}$,
although our computational domain does not cover
the X-ray emitting region.
We need the numerical simulations with 
larger computational domains.

The density and the temperature of the SS433 jets 
near the compact object are investigated by 
\citet{Inoue01}.
They analytically studied 
and reported $\rho \sim 7\times 10^{-7} \rm g \, cm^{-3}$
at $r\sim 9\times 10^7 \rm cm$.
Since this size corresponds to $r\sim 2\times 10^2 r_{\rm S}$
for a mass of $1.4M_\odot$,
their result roughly agrees with our result (see the top panel).
On the other hand, 
whereas we used the one-temperature assumption 
in the present simulations,
it is not guaranteed in the outflow regions.
This is because 
the equipartition time, $\lsim 10^{-3}$ s,
is comparable to, or larger than, the escape time
at the regions of $r\lsim 200r_{\rm S}$.
In the case in which the one-temperature assumption breaks down,
the ion temperature would be 
much higher than the electron temperature.
Such a tendency was also reported by \citet{Inoue01}.
In their work, the ion temperature
was one order of magnitude
higher than the electron temperature.
However, their model predicts a much higher 
electron temperature, $\sim 2\times 10^{10} \rm K$,
than the result of our model, $\sim 10^9 \rm K$.
Although we solved a set of RHD equations
in the present simulations,
\citet{Inoue01} investigated the only the energetics
by assuming the dynamics and the shape of the jets.
We should perform RHD simulations, in which
the ion and electron temperatures are separately treated.

\subsection{Future work}
\label{Future}
We need to calculate the spectra and compare them with 
the results of the observations for detailed study.
A symptom of the supercritical accretion onto NSs
has been discovered by the X-ray observations.
\citet{TM06} reported that 
the LMXB spectra can be understood by the superposition
of emissions from an accretion disk, the NS surface, and outflows.
Since the emission, which is thought to come from the outflows,
is not detected in the subcritical phase,
they concluded that the outflows are accelerated by
the strong radiation force of the supercritical accretion flows.
By solving the monochromatic radiation-transfer equation,
we calculated the effective-temperature profiles,
and show them in Figure \ref{Teff}.
However, the emergent spectra of the supercritical accretion flows
are not a simple superposition of blackbody spectra
with various effective temperatures \citep{OMW03}.
In addition, the inverse Compton scattering would also 
contribute to the spectra,
though it is not included in our simulations.
A detailed study of the emergent spectra should be
explored in future work.

In the present study, 
we consider the accretion flows 
around non-rotating NSs.
Our results do not change that much 
as long as the rotation velocity of the NS surface is
much smaller than the Keplerian velocity.
Here, we note that the spin-up timescale of the NSs 
by mass accretion, $\sim 10^8$ yr,
is much longer than the integration time of the present simulation,
$\sim 10$ s.
The rapidly rotating NSs seem to exist in the LMXBs,
since they are thought to be progenitors of millisecond
radio pulsars 
\citep{Backer82, Wijnands98}.
The rapid rotation plays important roles 
in the disk structure and the temperature profile
via modification of the space-time geometry
(\authorcite{Bha00} \yearcite{Bha00}, 2001a, b).
A detailed consideration of the disks around the rapidly rotating NSs,
however, beyond the scope of this paper.

We consider the non-magnetized NSs in the present study.
If the NSs have strong magnetic fields,
the flow structure might drastically change
\citep{Kato01, Koldoba02, Romanova03}.
In addition,
the source of the disk viscosity is likely to be of magnetic origin,
though we used the $\alpha$-prescription of the viscosity
\citep[for a review]{HBS01, MMM01, Balbus03}.
We need radiation magnetohydrodynamic (MHD) simulations.
Local radiation MHD simulations of the 
accretion flow have been performed by 
\citet{Turner03} and \citet{Turner04}.
Furthermore, 
we should include the synchrotron self-Compton processes.
It is important to separately 
treat the ion temperature and the electron temperature.
As we discussed in \S \ref{D_SS433},
the one-temperature assumption might break down in the outflow regions
because of low density and high velocity.
The radiation MHD simulations are required again for such a study.

Although the matter did not lose angular momentum
by the gas-radiation interaction in the present study,
the radiation drag would extract angular momentum from the disk.
However, we here stress that our results do not change very much
even if we take the radiation drag into consideration.
\citet{ML93} have reported that the mass accretion is enhanced 
via radiation drag by the radiation from the NSs.
The radiation drag-induced mass-accretion rate is given by
$L_{\rm NS}/c^2$, if all photons emitted at the NS surface
contribute to extract the angular momentum
(see also \authorcite{Umemura01} \yearcite{Umemura01}).
This value is much smaller than the resulting mass-accretion rate
(see Figure \ref{Lum}).
Additionally, the radiation drag-induced mass-accretion rate 
by the radiation from the disk 
is $L_{\rm rad}/c^2$ maximum.
It is also smaller than our results as shown in Figure \ref{Lum}.
Since photons are emitted from rapidly rotating matter,
and since some part of the photons escape through the rarefied region
above the disk without any interaction with the disk matter,
the efficiency of the radiation drag would decrease.
Thus, we conclude that our present results do not change very much
due to radiation drag.
For a detailed study,
we should self-consistently solve the 
radiation fields and the accretion flow,
by performing the relativistic RHD simulations.

\section{CONCLUSIONS}
By performing two-dimensional RHD simulations, 
we investigated the supercritical accretion flows 
around the NSs. 
We summarize our results as follows.

(1)
Since the radiation energy, as well as the gas energy, is not 
swallowed by the NS,
and since the rotation energy is effectively converted to
radiation energy at the innermost regions via the viscosity,
the radiation force is enhanced near the NS surface. 
The enhanced radiation force counteracts with the gravity
and produces high-density regions around the NSs.
Such a high-density shell is a peculiar structure 
of supercritical accretion flows onto the NSs,
although the disk and outflows are also produced 
in the case of the BH accretion flows.

(2)
Although the enhanced radiation force reduces 
the mass-accretion rate, and drives the strong outflows
above and below the disk,
supercritical disk accretion onto the NS is possible. 
The mass-accretion rate onto the NS 
is $20-30\%$ of that onto the BH,
on the condition that we employ the same mass-input rate.
The mass-outflow rate is a few-times larger 
in the NS accretion flows than in the BH accretion flows.

(3)
The accreting matter releases 
most of its energy before reaching the NS surface.
Due to this mechanism,
the luminosity of the NS is insensitive to the mass-accretion rate.
Therefore, although the matter accretes onto the NS at
the supercritical rate,
the radiation force does not exceed the gravity
at the NS surface.

(4)
The supercritical accretion flows onto the NSs mainly release
the accretion energy as the kinetic energy of the outflows,
especially for a large mass-accretion rate.
In contrast, the kinetic energy output rate via the outflows
does not exceed the disk luminosity 
in the case of BH accretion flows.


(5)
Our simulations succeed in reproducing 
the observed properties of the SS433 jets,
except for the collimation angle.
The resulting jet velocity ($0.2-0.3c$)
and mass-outflow rate ($150-700L_{\rm E}/c^2$)
agree with the observations.
However, the collimation angle of the jet in our simulations
($\sim 20^\circ$)
is larger than that of SS433 (a few degrees).

\bigskip
\bigskip

We would like to thank the an anonymous reviewer for many helpful comments.
We especially thank N. Shibazaki, H. Susa, H. Takahashi, S. Mineshige,
Y. Kato, and T. Harada for useful comments and discussions.
The calculations were carried out 
by a parallel computer at Rikkyo University
and Institute of Natural Science, Senshu University.
This work is supported in part by a special postdoctoral
researchers program in RIKEN.
We acknowledge a research grant from Japan Society 
for the Promotion of Science (17740111).

\end{document}